\documentstyle[12pt,aaspp4,psfig]{article}

\input epsf.sty

\begin{document}


\title{Contribution of Bright Extragalactic Radio Sources to
      \\Microwave Anisotropy}

\author{Aaron Sokasian
\footnote{Current address:  Harvard-Smithsonian
Center for Astrophysics, 60 Garden St., 
Cambridge, MA  02138}$^,$\footnote{asokasia@cfa0.harvard.edu}
}
\affil{Department of Astronomy, Cornell University, 
Ithaca, NY  14853
}
\author{Eric Gawiser\footnote{gawiser@astron.berkeley.edu} and 
George F. Smoot\footnote{smoot@cosmos.lbl.gov}}
\affil{Department of Physics, University of California, Berkeley, 
and Lawrence Berkeley National Laboratory, Berkeley, CA  94720}
\authoremail{asokasia@cfa0.harvard.edu}

\begin{center}
 submitted to the {\it Astrophysical Journal}
\end{center}

\begin{abstract}

We estimate the contribution of extragalactic radio 
sources to fluctuations in  sky temperature 
over the range of frequencies 
(10-300 GHz) used for Cosmic Microwave Background (CMB) 
anisotropy measurements.  CMB anisotropy observations at 
high resolution and low frequencies are especially sensitive 
to this foreground.  We have compiled a catalog of 2207 bright radio 
sources, including 758 sources with flux measurements at 90 GHz.  
We develop a method to extrapolate the source spectra and  
predict skymaps of extragalactic radio sources at instrument 
resolutions of $10'-10^{\circ}$ FWHM.  
Our results indicate that the brightest radio sources will dominate microwave
anisotropy for a wide range of resolutions and frequencies.  
Our skymaps predict the 
location and flux of the brightest radio sources at each frequency, 
making it straightforward to develop a template for masking 
the pixels containing them.  This masking should be sufficient to protect
high resolution CMB anisotropy observations from unacceptable radio 
source confusion.

\end{abstract}


\section{Introduction}

The Cosmic Microwave Background Explorer (COBE) detection 
of large angle CMB anisotropy (Smoot et al. 1992) has sparked a drive 
to measure the anisotropy on smaller angular scales with the goal 
of determining crucial information about the density and 
expansion rate of the universe, the 
nature of dark matter, 
and the spectrum of primordial density perturbations.  
COBE DMR observations 
were basically unaffected by extragalactic foreground sources 
(Banday et al. 1996, Kogut et al. 1994) due to the large beam size 
($7^{ \circ}$ FWHM).
Because the contribution of a point source increases with the 
inverse of the beam area, observations at higher 
angular resolution are more sensitive to extragalactic foregrounds, 
including radio galaxies, bright infrared 
galaxies (Gawiser \& Smoot 1997), high-redshift infrared galaxies 
(Blain et al. 1998, Gawiser, Jaffe, \& Silk 1998, hereafter GJS),
and the Sunyaev-Zeldovich effect 
from galaxy clusters (Aghanim et al. 1997). 
Estimates of extragalactic foreground confusion 
are critical as many ground-based, balloon-borne, and 
satellite experiments (MAP, Planck Surveyor) 
plan to study CMB anisotropies at angular scales
from 5$`$ to $30`$, and preliminary results are already
available (e.g. Netterfield et al. 1997, Scott et al. 1996).  

To evaluate the impact of known 
radio sources on CMB anisotropy observations, 
we use flux data from a variety of catalogs (see $\S$2),
including recent measurements, 
to construct models of source spectra as a function of frequency.  
We analyze simulated skymaps at frequencies from 
10 to 300 GHz to determine the expected contribution of 
radio galaxies to foreground 
confusion of CMB temperature anisotropy.  
This information will be useful when choosing frequencies and 
regions of the sky to observe CMB fluctuations on small angular scales.
This work represents a significant improvement over previous efforts 
(Toffolatti et al. 1995, 1998, Franceschini et al. 1989)
which depended upon galactic evolution models
to predict the contribution of simulated 
radio sources at microwave frequencies.  Our
catalog contains detailed observations of known sources 
and hence can be used to make a spatial template 
for masking out their emission, and we believe that this 
phenomenological approach will lead to greater accuracy in predicting source
counts and the overall level of foreground anisotropy.

\section{Our Catalog}

	The discrete radio sources used in this project were compiled 
from a number of separate catalogs. Our current catalog includes flux 
measurements and their corresponding errors at multiple frequencies for 
2207 sources.  
We have focused our attention on obtaining all 
available radio observations at millimeter and sub-millimeter wavelengths, 
resulting in 5766 observations of 758 different sources at 90 GHz, 
890 observations of 229 different sources from 100-200 GHz, and 
2628 observations of 309 different sources at frequencies above 200 GHz.  
The sources are roughly isotropic in distribution, 
except for a significantly greater number of sources 
in the northern celestial hemisphere 
due to the anisotropic distribution of radio telescopes on Earth.  
In addition, there are noticeably fewer observations 
within $10^{\circ}$ of the 
galactic plane and the celestial north pole due to the difficulty of 
observing extragalactic radio sources in those locations.  
  
Our catalog includes the full-sky 5 GHz-selected 
1 Jy sample of K\"{u}hr et al. (1981).  We add  
high-frequency ($>90$ GHz) 
measurements (Steppe et al. 1988, 1992, 1995; 
Tornikoski et al. 1996;
Kreysa 1998;
Antonucci, Barvainis, \& Alloin 1990;
Beichman et al. 1981;
Chini et al. 1989;
Gear et al. 1994;
Edelson 1987;
Holdaway, Owen, \& Rupen 1994;
Knapp \& Patten 1991;
Landau et al. 1980, 1983, 1986;
Lawrence et al. 1991;
Nartallo et al. 1997; 
Owen et al. 1978;
Owen, Spangler, \& Cotton 1980;
Stevens, Robson, \& Holland 1996; 
Chandler 1995;
VLA Memo\#193) 
and centimeter-wavelength observations  
(Herbig and Readhead 1992; Patnaik 1992; 
Wiren et al. 1992;
Stanghellini et al. 1997; 
Perley 1982; Cotton 1980;
Aller et al. 1985;  
VLA Calibrator Manual).  An updated version of the catalog with 
more detail will be described by Gawiser, Sokasian, \& Smoot (1998, 
hereafter GSS).  

\section{Spectral Fitting}
Some 
extragalactic radio sources have complex spectra which cannot be 
approximated by simple functional forms due to emission from 
both compact and extended structures which dominate at different 
frequencies.  
In most radio galaxies, 
the emission comes from radio lobes located symmetrically around the core.  
The dominant emission mechanism, synchrotron, 
can be well approximated by a simple power law,
\begin {equation}
S\propto \nu^{-\alpha}
\end {equation}

\noindent with a flux spectral index, $\alpha$, typically between 0.5 and 
1.0 (Platania et al. 1997).  
Some radio sources have 
compact active nuclei which generate flat-spectrum radio emission.   
The spectra of these sources can be  
inverted ($\alpha < 0$) for most of the 
radio frequency range due to self-absorption of the lower frequency emission.
Attempts to fit the observational 
data have yielded a variety of results.
The central engine of a typical active 
galaxy may consist of a supermassive black hole surrounded by an accretion  
disk and accelerating a jet of relativistic particles perpendicular to the 
disk plane (e.g. Urry \& Padovani 1991). 
 B\"{o}ttcher et al. (1997) proposed 
a  model in which the inverted spectrum  of NGC 3031 is assumed to be the 
emission of a jet component, becoming optically thin to the radio emission 
of a monoenergetic pair plasma at decreasing frequencies as it moves outward 
and expands.  

For sources which lack direct high-frequency observations, we 
avoid trying to determine the nature of the emission mechanism.  Instead, 
we use a phenomenological approach based on the expectation that the 
spectra of most radio sources approach power law behavior at 
frequencies higher than $\simeq 5$ GHz 
(Verschuur \& Kellerman 1988).
This power law may then be used to extrapolate the spectrum  
to typical CMB observation frequencies.  
A previous phenomenological approach (Tegmark \& Efstathiou 1996) 
extrapolated 1.4 GHz source counts by assuming flat-spectrum emission for 
all sources. 
Our method has the advantages of using 
the actual source locations,
which can be turned into templates 
for masking the brightest pixels on the sky.  
To determine if there is any spectral index clustering, 
we use sources that have been measured near 1.4, 10, and 90 GHz and 
plot each source's spectral index from 1.4 GHz to 10 GHz versus its index
from 10 GHz to 90 GHz in Figure 1.  
There is a vague clustering of bright sources (circles)  
consistent with the notion that 
the brightest sources selected at low frequencies  
tend to have steep spectra.  The overall scatter of source spectra 
in Figure 1 
shows
that it is wrong to categorize radio sources into template spectra
or a narrow spectral index range.  This motivates us to  
 fit the spectra of each source individually.

To determine the frequency beyond which a power law (a line on a 
log-log plot) can be fitted to the 
spectra of a given source, we use an iterative model 
which starts with the best-fit line to the three highest frequency data points
and repeatedly includes the next highest frequency data point to the set to
which it fits a line.  
The fitting stops when the reduced $\chi^{2}$ 
starts to get worse or becomes acceptable ($\simeq 1$).
 There is little evidence that inverted spectra are common
past 30 GHz (Steppe et al. 1995, 
Stanghellini et al. 1997), so 
we set the few inverted ($\alpha \geq 0$)
 high-frequency spectra in the 
catalog to flat ($\alpha = 0$) spectra.  These inverted spectra appear 
to result from variable sources being observed at different epochs 
at different frequencies, and we find that most of the sources 
with $\alpha_2 \leq 0$ in Figure 1 based on their mean 10 and 90 
GHz fluxes are better fit by an $\alpha \geq 0$ power-law when 
all observations are taken into account.  
The average high-frequency spectral index was 0.5
with 27\%  of the sources in our catalog 
having steep spectra ($\alpha > 0.75$), 
and 37\% having flat spectra ($\alpha < 0.25$).

To check the accuracy of this technique, we ran our extrapolation 
method on sources with observations at 90, 150, and 230 GHz  
while ignoring the observations above  
certain frequencies and then compared the measured fluxes  
with the extrapolated fluxes.  
The results (Table 1) show that the 
extrapolation method works best when there is at least one measurement 
at 20 GHz or greater, as expected since many spectra become power-law 
past 5 GHz.  Table 1 shows that on average we overpredict the 
flux at 90 GHz by a factor of 1.6, even when measurements above 20 GHz 
are used.  However, the median such error factor is only a factor 
of 1.1 overestimate, so we have roughly an equal number of over- and 
under-estimates.  This is no longer the case at 230 GHz, where even 
the median error factor is 1.9; our extrapolation method is overestimating 
the typical flux due to flat spectra 
falling off to more typical synchrotron spectra at frequencies around 100 GHz
(Gear et al. 1994).
It is difficult to predict how far this fall-off will last, as thermal 
emission from low levels of dust in these radio-bright galaxies are 
expected to dominate their spectra by 500 GHz, except for the BL Lacs 
which have flat spectra up to infrared wavelengths (Knapp \& Patten 1991, 
Chini et al. 1989, Landau et al. 1986).
  We therefore 
only trust our extrapolation in the range that has been tested, up to 
a maximum frequency of 300 GHz.  As the radio sources that have been 
observed at 30-300 GHz were selected at lower frequencies for 
brightness and flat spectra, 
our errors are only good estimates for this type of radio sources.  
This selection effect is not a great concern, however, as 
those are exactly the type of radio sources that threaten CMB 
anisotropy observations.
When interpolation is required, we use a cubic spline which passes 
through the mean fluxes at the observed frequencies.  
We visually inspected all 2207 sources to check the algorithm and 
eliminate any serious errors or outliers.  

For planned CMB anisotropy experiments, an additional concern is that 
the flat-spectrum radio sources can vary by up to a factor of ten 
in flux since their 
emission comes from a compact, active core.  Typical variations occur 
on timescales of one month to one year, although the overall 
spectrum shape is often preserved for a decade or longer 
(Tornikoski et al. 1993).   
We use the scatter in 
the observed fluxes of a source at each frequency to estimate the typical 
range of variability, which yields an error bar on the source's flux 
at that frequency about the mean of all observations.  Because the 
variations are not periodic, there is little more that can be done, unless 
sources are observed nearly simultaneously at higher resolution and nearby 
frequencies.  GSS looks at the issue of variability 
in detail, including the possibility of extrapolating long-term drifts 
in source flux to the next epoch of observation.  Radio sources 
are typically 4-7\% polarized,
and this polarization is variable (Nartallo et al. 1997), so 
radio-source foreground subtraction will be an important consideration
 for CMB polarization 
observations as well.

\section{Results}

We use the fitted spectra to predict the 
microwave flux of each radio galaxy in Jy ($1$ Jy $= 10^{-26}$ W/m$^2$/Hz).  
To convert from flux $S$ to antenna temperature $T_A$, we use

\begin{equation}
 T_A = S \frac{\lambda^2}{2 k \Omega}\; \;,  
\end{equation}

\noindent 
where $k$ is Boltzmann's constant, $\lambda$ is the wavelength, and 
$\Omega$
is the effective beam size of the observing instrument.  The 
antenna temperature of the Cosmic Microwave Background radiation, which 
has a thermodynamic temperature of T$=2.73$K, is given at frequency $\nu$ by

\begin{equation}
 T_A =  \frac {x}{e^x - 1} \; \; T,  
\end{equation}

\noindent
defining $x \equiv h \nu / k T$.  Fluctuations in antenna 
temperature caused by point sources will appear as 
thermodynamic temperature fluctuations in the CMB according to

\begin{equation}
 \frac{dT}{dT_A} = \frac {(e^x - 1)^2} { x^2 e^x} \; \; .  
\end{equation}
\noindent
The intrinsic $\Delta T / T$ of the CMB found by COBE is $\simeq 10^{-5}$
and is expected to vary between that and $3 \times 10^{-5}$ at the 
angular resolutions considered here.

An analysis of source counts indicates that the northern celestial 
hemisphere subset of our catalog 
appears to be complete down to an extrapolated flux of  1.0 Jy at 
90 GHz while the southern hemisphere is incomplete 
below 2.0 Jy at 90 GHz.  
For the purposes of statistical analysis we have concentrated on the 
northern hemisphere where we appear to have measurements of the 
200 
brightest sources in the hemisphere.  
We cannot rule out the existence of an unrelated
population of sources peaking around 90 GHz which are not bright at lower 
frequencies, as 90 GHz observations have only been made for 
sources selected at frequencies below 10 GHz (this hypothetical
source population is limited by GJS).  The brightest
sources will dominate the anisotropy unless they are masked, 
because uncertainty in their exact fluxes makes subtraction highly 
inaccurate.  After masking, 
the brightest remaining sources will dominate unless non-Poissonian 
clustering becomes appreciable.  Toffolatti et al. (1998) 
have shown that
non-Poissonian clustering is not expected to make an important contribution
to the foreground anisotropy from radio sources.  

To simulate observations, we convolve  all sources on pixelized sky maps 
(twice oversampled) of resolution varying from 
$10'$ to $10^{\circ}$ at frequencies between 
10 and 300 GHz.  
The information contained in these sky maps can be used to choose 
regions for observation (Smoot 1995) and pixels
to be masked during data analysis.  Figure 2 shows a 30$'$ resolution 
skymap at 100 GHz; pixels at the maximum of the color table should
be masked.  To avoid underestimating the anisotropy and to reduce the 
possibility of residual galactic contamination, we use only the 
portion of each skymap which 
covers galactic latitudes $|b| > 30^{\circ}$ and corresponds to the northern
celestial hemisphere to produce estimates of $\Delta T / T$.  

Figure 3 shows a summary of our results for several relevant 
instrument resolutions.  
The inverse relationship between anisotropy and 
FWHM arises due to the combined effects of beam convolving and pixelization.  
The exact level of oversampling causes a small change
in the measured anisotropy, but the 1/FWHM behavior should hold for
extrapolation to smaller resolutions (see GJS).  

We also analyze $\frac{\Delta T}{T}$ in the northern hemisphere 
based on only the 758 sources with 90 GHz measurements.  
The resulting rms 
$\frac{\Delta T}{T}$ at 90 GHz with a FWHM of $30'$ 
is $2\times 10^{-6}$ which dominates the anisotropy since the  
rms $\frac{\Delta T}{T}$ from extrapolating the spectra of 
the other 1449 sources 
amounts to only  $7\times 10^{-7}$.  
This indicates 
that we have flux measurements for the vast majority of bright 90
GHz radio sources.  
Refregier, Spergel, \& Herbig (1998) 
find that the $5 \sigma$ source detection limit for 
$0\fdg3$ MAP pixels will be 2 Jy at 90 GHz.  We have 108 sources 
in our catalog which have been observed to be brighter than 2 Jy at 90 GHz 
at least once, but only 42 sources have a weighted average flux that high, 
and a total of 52 sources are predicted to be brighter than 2 Jy at 90 GHz.  
We estimate that there will be 40-50 radio sources on the sky 
brighter than 2 Jy at 90 GHz.  At the 0.4 Jy level, 
Toffolatti et al. (1998) 
predict roughly twice as many sources as we do, but our 
prediction falls within their range of uncertainty.  
As the source counts we predict at this level based on the 
northern celestial hemisphere should be nearly complete, 
we recommend a slight recalibration of the galaxy evolution models 
used by Toffolatti et al., 
although a factor of two represents remarkable agreement 
for such different approaches.  

Table 2 lists the expected level of anisotropy and the 
number of detected radio sources in MAP and Planck channels 
if this type of straightforward $5 \sigma$ source detection and subtraction 
is performed.  
Since the 90 GHz MAP channel will have a resolution close to $0\fdg2$ we 
expect a $5 \sigma$ source detection limit of 1 Jy at 
90 GHz.  
These detected sources represent a list of the few hundred 
brightest radio sources in the sky at each frequency.  
The anisotropy levels are shown in Figure 3.  
The level of source confusion drops if the 
brightest sources ($\geq 5 \sigma$) are subtracted. 
Table 3 shows how the expected level of temperature 
anisotropy from radio sources varies with cutoff level, where we use 
our catalog to obtain prior information on which pixels are expected 
to contain sources at a given flux level and then mask those pixels.  
While all $5 \sigma$ pixels can be masked without such prior information 
if the CMB anisotropies are assumed to follow a Gaussian distribution, 
it is impossible to remove all $1 \sigma$ pixels without 
crippling the 
analysis.  
The actual improvements 
from masking all pixels expected to contain $1 \sigma$ sources may be 
less than indicated, unfortunately, due to the effect of incompleteness 
in our catalog.  GSS attempt to fill in 
this incompleteness using full-sky catalogs at 5 GHz.  If we were to settle
for making the southern celestial hemisphere as complete as the northern 
is now, we could create a mask for all sources expected to contribute 
at the 3$\sigma$ level, which is enough to make a significant reduction 
in radio source contamination versus 5$\sigma$ subtraction alone.

\section{Discussion}
Spectral analysis of bright radio sources 
indicates that their spectra are complex and cannot in general 
be categorized into template spectra or single power-laws.
The results from our analysis of extrapolation errors suggest that our 
phenomenological approach of fitting a power law to the 
high-frequency end of each spectrum is a reasonable model to use to 
extrapolate radio sources to microwave frequencies.  
Although subject to systematic errors on a galaxy-by-galaxy basis, 
we expect our overall extrapolation results to be accurate to within 
a factor of two at 90 GHz.

Our analysis of foreground 
confusion from extragalactic radio sources 
indicates that they contribute negligibly to COBE resolution 
observations of the CMB, consistent with the conclusion of 
Banday et al. (1996).  
However, they do become problematic at higher resolution.  
Our results set a lower limit on the anisotropy and provide a 
list of the brightest sources in the sky which can be used to mask pixels in 
future high-resolution CMB observations.
The contribution of extragalactic radio sources to CMB anisotropy is comparable
at 200 GHz to that of bright extragalactic infrared sources 
(Gawiser \& Smoot 1997, 
Toffolatti et al. 1998).  
Our current
results indicate a valley at around 200 GHz where the anisotropy from
radio sources is a minimum; adding in the contribution from infrared-bright
galaxies should move that valley towards 150 GHz.  

The results of this investigation motivate an expansion of our catalog 
so that sources which will 
contribute to 
anisotropies on the 
1$\sigma$ level can be masked.  
It is clear that the current generation of 
CMB anisotropy experiments must pay close attention to the possibility of 
radio point source contamination at all frequencies.  Masking
pixels which contain bright radio galaxies should reduce this foreground
to a manageable level.

\section {Acknowledgments}

We thank Cameron Murray and Chris Witebsky for their diligent work in 
compiling the radio catalog, 
Jon Aymon for writing IDL display routines,  
Tom Herbig and Chris Witebsky 
for providing us with useful software, 
Ernst Kreysa, Harri Terasranta, Merja Tornikoski, and Esko Valtaoja 
for providing us with recent data pre-publication, 
Mark Holdaway
for helping us obtain the MMA Memo, and Bruce Partridge, 
Malcolm Bremer, David Spergel, Alex 
Refregier, Alastair Edge, Andrew Blain, Jason Stevens, 
Doug Finkbeiner, and Jose Luis Sanz for helpful conversations.    
A.S. acknowledges the support of an LBNL Summer Research Fellowship.
E.G. acknowledges the support of an NSF Graduate Fellowship.  This 
work is supported in part by the DOE Contract No. DE-ACO3-76SF00098 
through Lawrence Berkeley National Laboratory and NASA Long Term 
Space Astrophysics Grant No. 014-97ltsa award \#NAG5-6552.  


\begin{references}

\reference{Aghanim et al. 1997}
Aghanim, N., De Luca, A., Bouchet, F. R., Gispert, G., \& Puget, J. L. 1997, 
\aap, 325, 9

\reference{Aller et al. 1985}
Aller, H. D., Aller, M. F., Latimer, G. E., \& Hodge, P. E. 1985, 
\apjs, 59, 513

\reference{Antonucci, Barvainis, \& Alloin 1990}
Antonucci, R., Barvainis, R., \& Alloin, D. 1990, \apj, 353, 416

\reference{Banday 1996}
 Banday, A. J., et al., \apj, 468, L85 

\reference{Beichman et al 1981}
 Beichman, C. A., Neugebauer, G., Soifer, B. T. Wootten, H. A., Roellig, T., 
\& Harvey, P. M. 1981, {\it Nature}, 293, 711

\reference{Blain, Ivison, Smail, \& Kneib 1998}
Blain, A.W., Ivison, R.J., Smail, I., \& Kneib, J.-P. 1998, 
 to appear in {\it Wide-field surveys in cosmology, Proc. XIV IAP meeting}, 
 astro-ph/9806063 

\reference{B\"{o}ttcher et al. 1997}
 B\"{o}ttcher, M., Reuter, H. P., \& Lesch, H. 1997, \aap, 326, 33, 
astro-ph/9707353

\reference{Chandler 1995}
Chandler 1995, VLA Memo \#192, unpublished

\reference{Chini et al 1989}
Chini, R., Biermann, P. L., Kreysa, E., \& Gemund, H. P. 1989, A\&A, 221, L3

\reference{Cotton 1980}
Cotton, W. D. 1980, \aj, 85, 351

\reference{Edelson 1987}
Edelson, P. A., 1987, \aj, 94, 1150

\reference{Franceschini et al. 1989} 
 Franceschini, A., Toffolatti, L., Danese, L., \& De Zotti, G. 1989, 
\apj, 344, 35  

\reference{Gear et al. 1994}
Gear, W. K. et al. 1994, \mnras, 267, 167

\reference{GJS}
Gawiser, E., Jaffe, A., \& Silk, J. 1998, astro-ph/9811148 [GJS]

\reference{Gawiser \& Smoot 1997}
Gawiser, E. \& Smoot, G. F., 1997, \apj, 480, L1

\reference{GSS}
Gawiser, E., Sokasian, A., \& Smoot, G. F., 
1998, in preparation [GSS]

\reference{Herbig \& Readhead 1992}
Herbig, T. \& Readhead, A. C. S., 1992, \apj, 81, 83

\reference{Holdaway, Owen \& Rupen 1994}
Holdaway, M. A., Owen, F. N. \& Rupen, M. P., 
1994, NRAO Report, MMA Memo \#123, unpublished

\reference{Knapp \& Patten 1991}
Knapp, G. R., \& Patten, B. M. 1991, \aj, 101, 1609

\reference{Kogut et al. 1996}
 Kogut, A., et al. 1994, \apj, 464, L5 

\reference{Kreysa 1998}
Kreysa, E. 1998, private communication

\reference{K\"{u}hr et al. 1981}
 K\"{u}hr, H., Witzel, A., Pauliny-Toth, I. I. K., \& Nauber, U. 1981, 
\aaps, 45, 367

\reference{Landau, Epstein, \& Rather 1980}
Landau, R., Epstein, E., \& Rather, J. D. G. 1980, \aj, 85, 363

\reference{Landau et al. 1986}
Landau, R. et al. 1986, \apj, 308, 78

\reference{Landau et al. 1983}
Landau, R. et al. 1983, \apj, 268, 68

\reference{Lawrence et al. 1991}
Lawrence, A. et al. 1991, \mnras, 248, 91

\reference{Nartallo et al. 1997}
Nartallo, R., Gear, W. K., Murray, A. G., Robson, E. I., \& Hough, J. H.
1997, \mnras, 297, 667, astro-ph/9712219 

\reference{Netterfield et al. 1997}
Netterfield, C. B. et al., 1997, \apj, 474, 47

\reference{Owen et al. 1978}
Owen, F. N., Porcas, R. W., Mufson, S. L., \& Moffett, T.J. 1978, \apj, 83, 685

\reference{Owen, Spangler, \& Cotton 1980}
Owen, F. N., Spangler, S. R., \& Cotton, W. D. 1980, \aj, 85, 351

\reference{Patnaik et al. 1992}
Patnaik, A. et al. 1992, \mnras, 254, 655  

\reference{Perley 1982}
Perley, R. A. 1982, \aj, 87, 859

\reference{Platania et al. 1997}
Platania, P. et al., 1997, \apj, 505, 473, astro-ph/9707252

\reference{Refregier, Spergel, \& Herbig 1998}
Refregier, A., Spergel, D. N., \& Herbig, T. 1998, astro-ph/9806349

\reference{Scott et al. 1996}
Scott, P. F. S. et al., 1996, \apj, 461, L1

\reference{Smoot et al. 1992}
Smoot, G. F. et al. 1992, \apj, 396, L1

\reference{Smoot et al. 1995}
Smoot, G. F. 1995, Astrophys. Lett. and Communication, 32, 297

\reference{Stanghellini et al. 1997}
Stanghellini, C., O'Dea, C. P., Baum, S. A., Dallacasa, D., Fanti, R.,
\& Fanti, C. 1997, \aap, 325, 943

\reference{Steppe et al. 1988}
Steppe, H., Salter, C. J., Chini, R., Kreysa, E., Brunswig, W., 
\& Lobato-Perez, J. 1988, \aaps, 75, 317

\reference{Steppe et al. 1992}
Steppe, H., Liechti, S., Mauersberger, R., Kompe, C., Brunswig, W., \& 
Ruiz-Moreno, M. 1992, \aaps, 96, 441 

\reference{Steppe et al 1995}
Steppe, H., Jeyakumar, S., Saikia, D. J., \& Salter, C. J. 
1995, A\&AS, 113, 409

\reference{Stevens et al 1998}
Stevens, J. A., Robson, E. I., \& Holland, W. S. 1996, \apj, 462, L23

\reference{Tegmark \& Efstathiou 1996}
Tegmark, M., \& Efstathiou, G. 1996, \mnras, 281, 1297

\reference{Toffolatti et al. 1997}
Toffolatti, L. et al. 1998, \mnras, 297, 117, astro-ph/9711085

\reference{Toffolatti et al. 1995}
Toffolatti, L. et al. 1995, Astro. Lett. and Communication, 32, 125

\reference{Tornikoski et al. 1996}
Tornikoski, M. et al. 1996, \aaps, 116, 157

\reference{Tornikoski et al. 1993}
Tornikoski, M., Valtaoja, E., Terasranta, H., Lainela, M., Bramwell, D., 
\& Botti, L. C. L. 1993, \aj, 105, 1680

\reference{Urry et al. 1991}
Urry, C. M. \& Padovani, P., 1991, \apj, 374, 431

\reference{Verschuur \& Kellerman 1988}
Verschuur, G. L., \& Kellerman, K. I., ed., Galactic and 
Extragalactic Radio Astronomy, 2nd ed., 1988, New York, Springer-Verlag, 
Chapters 13,15

\reference{VLA}
 VLA Calibrator Manual, 1997, unpublished

\reference{VLA193}
VLA Memo \#193, unpublished

\reference{Wiren et al. 1992}
Wiren, S., Valtaoja, E., Terasranta, H., \& Kotilainen, J. 1992, \aj, 
104, 1009

\end {references}

\clearpage
\begin{table}[htb]
\caption{
Average Errors from Extrapolation.  The average 
extrapolation error is the mean of $\mid (S_P - S_O) / S_P \mid$ where 
S$_P$ is the predicted flux and S$_O$ is the observation.  
The average error factor is the mean of $S_P/S_0$.  
}

\begin{center}
\begin{tabular} {l|l|l|l}

\hline
Frequency tested & Frequencies ignored & 
Avg. Extrapolation Error & 
Avg. Error Factor \\
\hline
90 GHz & $\geq$ 2 GHz & 209 \% & 2.5  \\
90 & $\geq$ 10 & 148 & 2.3  \\
90 & $\geq$ 20 & 133 & 2.0  \\
90 & $\geq$ 90 & 92 & 1.6  \\
150 & $\geq$ 90 & 94 & 1.5 \\
230 & $\geq$ 90 & 250 & 3.2 \\
\hline
\end{tabular}
\end{center}
\end{table}

\begin{table}[htb]
\caption{
Foreground Anisotropy in MAP \& Planck channels after source removal.  
Sources which contribute to the anisotropies at the $5 \sigma$ level 
or higher are considered detected and can be 
removed by masking the pixels containing them.  No attempt 
has been made to use multi-frequency information or further prior 
information to detect and remove dimmer sources.  
}
\begin{center}
\begin{tabular} {r|c|l|c|l}

\hline
Frequency (GHz) & FWHM  & 
Source detection limit  &
\# detected &  
$\Delta T/T$ \\
\hline
MAP 20 & 56$'$ & 1.4 Jy & 186 & $8\times10^{-6}$     \\
30 & 41 & 1.2 & 216 & $4\times10^{-6}$   \\
40 & 28 & 0.9 & 265 & $3\times10^{-6}$  \\
60 & 21 & 1.1 & 168  & $2\times10^{-6}$  \\
90 & 13 & 1.0 & 161  &  $1.5\times10^{-6}$  \\
Planck 30 & 33 & 0.9 & 290 & $5\times10^{-6}$     \\
44 & 23 & 0.8 & 285 & $3\times10^{-6}$   \\
70 & 14 & 0.6 & 360 & $2\times10^{-6}$   \\
100 & 10 & 0.6 & 304  &$1.3\times10^{-6}$   \\
143 & 7 & 0.6 &  323  &$9\times10^{-7}$    \\
217 & 5  & 0.3 & 533 & $7\times10^{-7}$     \\
353 & 4.5 & 0.2 & 644 &$9\times10^{-7}$    \\
545 & 4.5 & 0.4 & 289 & $8\times10^{-6}$   \\
857 & 4.5 & 0.7 & 125  & $4\times10^{-4}$  \\
\hline
\end{tabular}
\end{center}
\end{table}

\begin{table}[htb]
\caption{Foreground Contamination in 13' MAP channel at 90 GHz.   
This analysis 
assumes that our catalog is used to identify 
sources whose fluxes will be above the threshold and that 
the pixels
containing those sources are masked.
The results given are 
for the northern celestial hemisphere, where our catalog is 
estimated to be complete for the brightest few hundred sources, so the 
final line is likely an underestimate of anisotropy.    
}
\begin{center}
\begin{tabular} {c|c|c}

\hline
Threshold (Jy)& 
\# Sources above Threshold & 
$\Delta T / T$ \\
\hline
None & 0 &  $4.4\times10^{-6}$     \\
2 (10 $\sigma$) & 49 &  $2.1\times10^{-6}$   \\
1 ($5 \sigma$) & 161 &  $1.7\times10^{-6}$   \\
0.6 ($3 \sigma$) & 346 &  $1.2\times10^{-6}$  \\
0.2 ($1 \sigma$) & 940 &  $3.8\times10^{-7}$  \\
\hline
\end{tabular}
\end{center}
\end{table}

\begin{figure}[htb]
\figurenum{1}
\centerline{\psfig{file=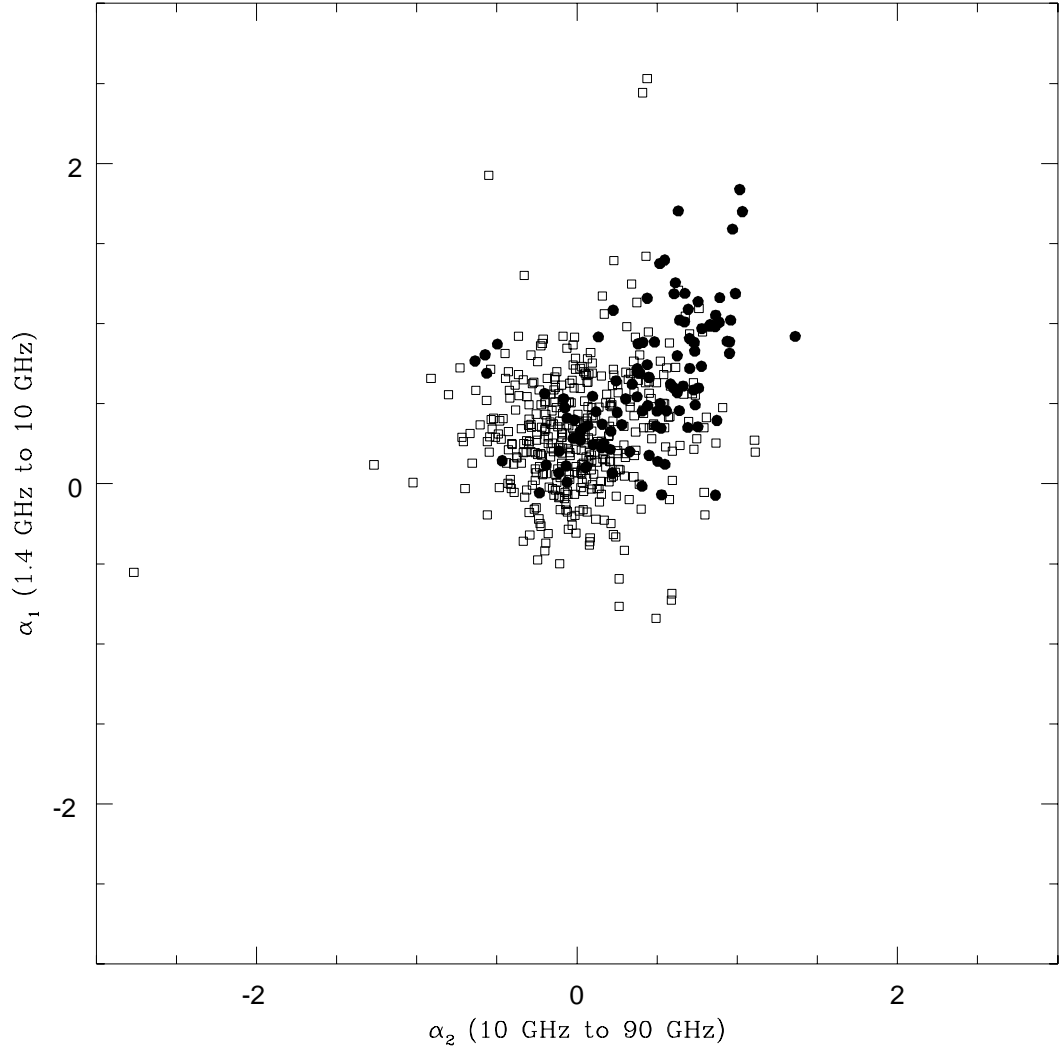,width=6in}}
\caption{
	Spectral indices $\alpha_1$ from 1.4 to 10 GHz and 
$\alpha_2$ from 10 to 90 GHz.   
Solid circles represent the brightest sources at 1.4 GHz; 
open squares represent dimmer sources at 1.4 GHz.  Note the lack
of clustering into distinct archetypal spectra.
}
\end{figure}

\begin{figure}[htb]
\figurenum{2}
\centerline{\psfig{file=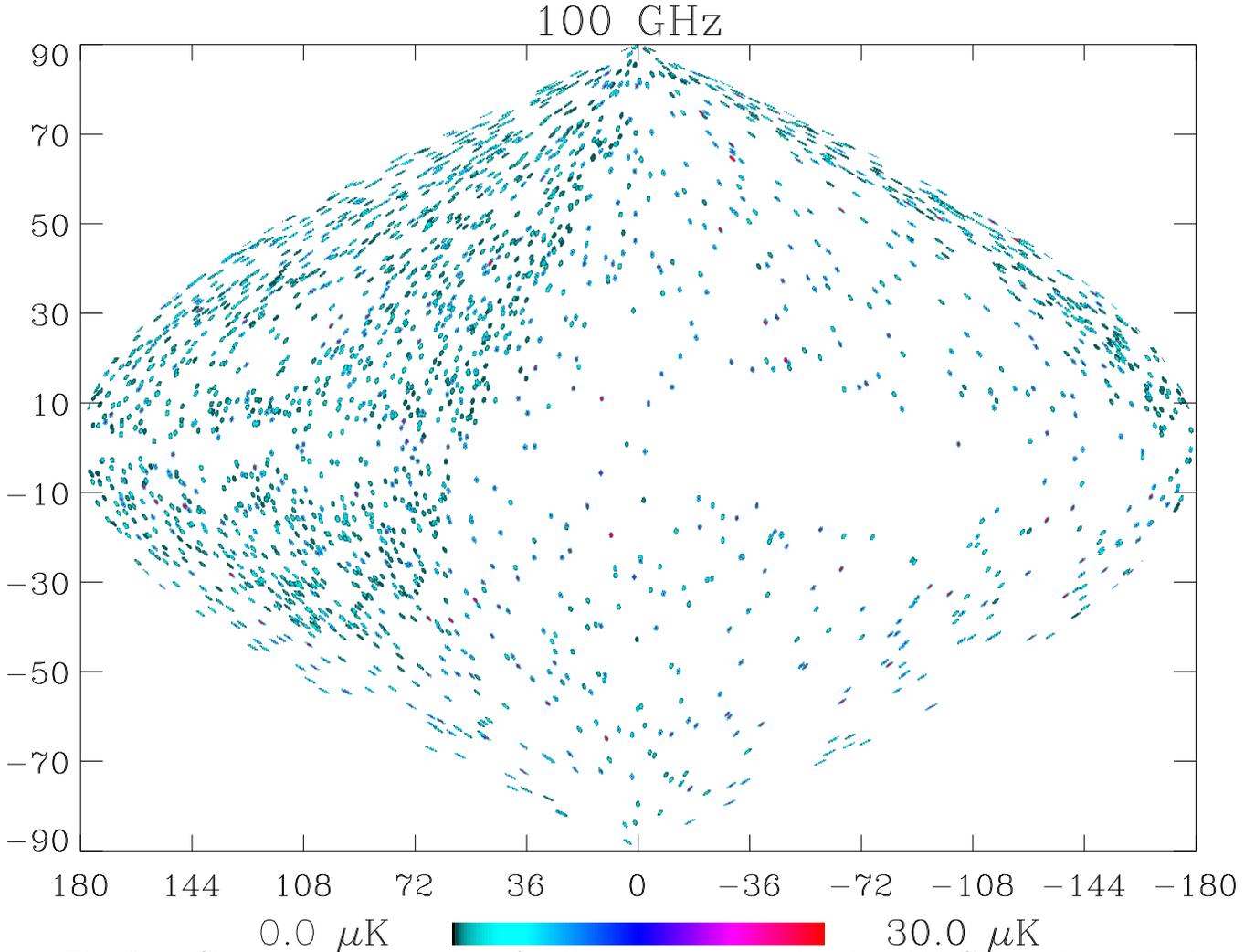,width=7.0in,angle=90}}
\caption{
Sky map of our catalog of radio sources extrapolated to 100 GHz and convolved
to simulate observation with a $0.5^{\circ}$ beam.  
The color table (thermodynamic 
temperature fluctuations) reaches a maximum
for sources which will be directly detectable by future satellites.
The plot is in Galactic coordinates, and the northern celestial 
hemisphere is seen to have a greater abundance of observed sources.
}
\end{figure}

\begin{figure}[htb]
\figurenum{3}
\centerline{\psfig{file=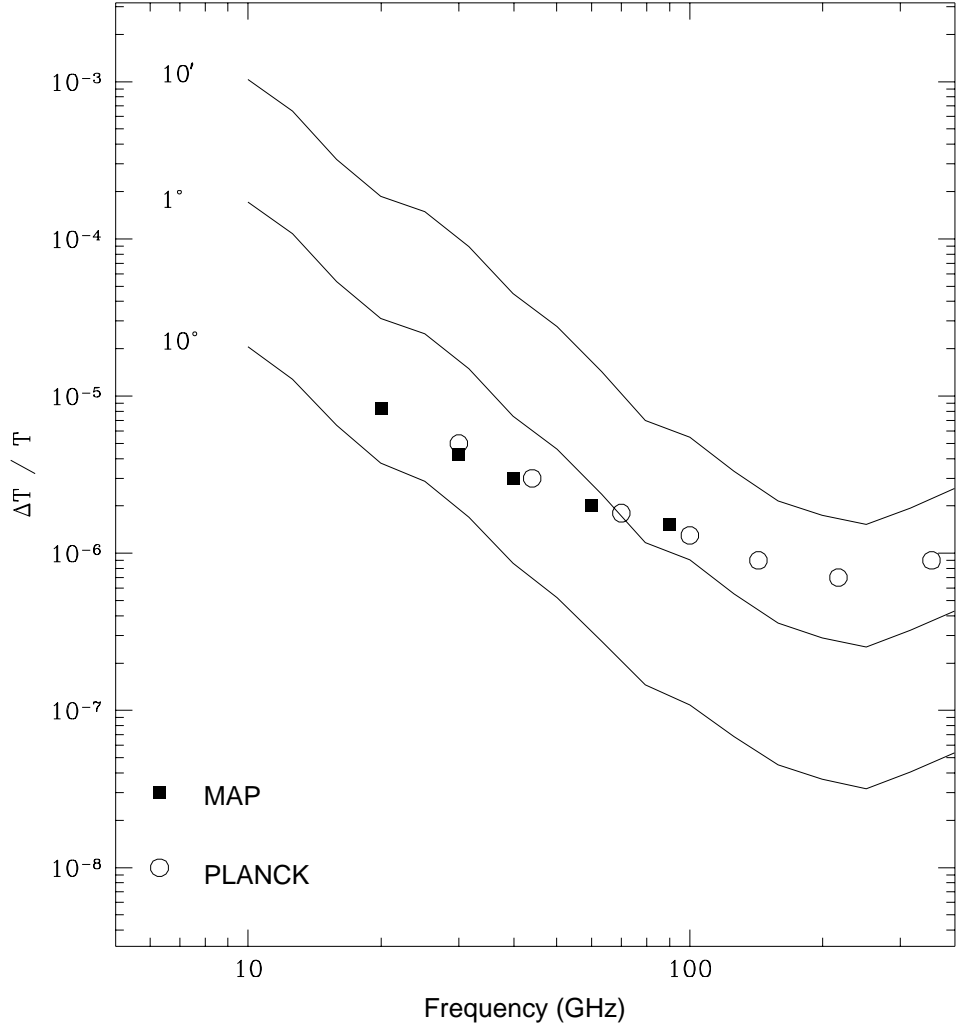,width=6in}}

\caption{
Plot of $\log_{10} \frac{\Delta T}{T}$ in the northern celestial 
hemisphere
(without pixel subtraction) versus frequency for instrument resolutions of
$10'$, $1^{\circ}$, and $10^{\circ}$,
showing window where foreground confusion should be
$\simeq 10^{-6}$.  
The rise beyond 200 GHz is 
caused by the exponential falloff in CMB antenna temperature beyond 100 GHz.  
The $5 \sigma$-source 
subtracted predictions for MAP (solid squares) and Planck (open 
circles) from Table 2 are also shown.  
The $1/FWHM$ scaling will extend to other instrument resolutions but 
the $5\sigma$ source detection threshold is mildly instrument dependent.
}
\end{figure}

\end{document}